\documentclass{piner}
\begin{document}

\title{The Fastest Relativistic Jets: VLBA Observations of Blazars
with Apparent Speeds Exceeding $25~c$}

\author{B.~G.~Piner\altaffilmark{1,2}, D. Bhattarai\altaffilmark{1},  
P.~G.~Edwards\altaffilmark{3}, and D.~L.~Jones\altaffilmark{2}}

\altaffiltext{1}{Department of Physics and Astronomy, Whittier College,
13406 E. Philadelphia Street, Whittier, CA 90608}

\altaffiltext{2}{Jet Propulsion Laboratory,
California Institute of Technology, 4800 Oak Grove Drive, Pasadena, CA
91109}

\altaffiltext{3}{Institute of Space and Astronautical Science, Yoshinodai, Sagamihara, Kanagawa 229-8510, Japan}

\begin{abstract}
We have measured peak apparent speeds of $25.6\pm7.0~c$, $25.6\pm4.4~c$, and $28.2\pm6.6~c$
in the jets of 0235+164, 0827+243, and 1406$-$076, respectively,
based on six epochs of high-sensitivity VLBA observations at 22 and 43 GHz during 2002 and 2003
($H_{0}$=71 km s$^{-1}$ Mpc$^{-1}$, $\Omega_{m}=0.27$, and $\Omega_{\Lambda}=0.73$).
These blazars had been identified as potentially having apparent speeds exceeding
40~$c$ in an earlier VLBA survey of EGRET blazars by Jorstad et al.
We therefore confirm (with high confidence in 0827+243, and lower confidence in 0235+164 and 1406$-$076)
the presence of highly relativistic pattern speeds in these three jets,
although not at the $>40~c$ levels reported by Jorstad et al.
The lower limit to the bulk Lorentz factor implied by the observed apparent speeds
is $\Gamma>\sim25-30$ in these three sources, if the pattern speeds are equal to or slower than the bulk flow speed.
\end{abstract}


\section{Introduction}
\label{intro}
Since the discovery of apparent superluminal motion in extragalactic jets with VLBI, there have
been occasional reports of very fast apparent speeds, exceeding 40~$c$ or so
(e.g., Jorstad et al. 2001; An et al. 2004).
Because the apparent speed of a jet places a lower limit on its bulk Lorentz factor
(under the assumption of equal bulk and pattern speeds), establishing the peak
apparent speed present in relativistic jets is equivalent to establishing the peak
Lorentz factor of the population.
The peak Lorentz factor is an important quantity for several reasons,
perhaps most importantly because
theoretical models of jet acceleration should be able to produce
jets that reach (but do not greatly surpass) this observed maximum Lorentz factor.

Some previous claims of very fast apparent speeds
have appeared upon subsequent observation
to most likely be a result of component misidentification among VLBI images
(e.g., 1156+295, McHardy et al. 1990; 
Piner \& Kingham 1997), a potential problem when comparing VLBI images
with possibly sparse $(u,v)$ plane coverage and a limited number of epochs.
Because of their importance, and because of the possibility of component misidentification,
candidate high-speed objects should be confirmed by follow-up observations.

In this paper, we report follow-up VLBI observations of the three sources
0235+164, 0827+243, and 1406$-$076,
reported by Jorstad et al. (2001) (hereafter J01) to have apparent speeds 
exceeding 40~$c$ (using our cosmological values, see below)
\footnote{The $>40~c$ apparent speed reported by J01 for 0827+243 was actually
due to an incorrect redshift for the source.
The speed of $40.2\pm7.6~c$ reported by J01 for 0827+243 (using our cosmological model)
is actually $22.7\pm4.3~c$ once the correct redshift is used.}. 
These three objects were observed by J01 during their EGRET blazar monitoring
program from 1993 to 1997.  Because they monitored 42 blazars, their integration
time on each and therefore their image dynamic range was limited, and some of their detections
of these fast components were only slightly above their noise levels.
We monitored these three sources with the
NRAO Very Long Baseline Array (VLBA)\footnote{The National Radio Astronomy Observatory is a facility of the National
Science Foundation operated under cooperative agreement by Associated Universities, Inc.},
using full-track observations to obtain images
of these jets with improved sensitivity over those in J01.
In this paper we use cosmological parameters
$H_{0}=71$ km s$^{-1}$ Mpc$^{-1}$, $\Omega_{m}=0.27$, and $\Omega_{\Lambda}=0.73$.
When results from other papers are quoted,
they have been converted to this cosmology.

\section{Observations}
\label{obs}
We observed 0235+164, 0827+243, and 1406$-$076 each at six epochs
with the VLBA between 2002 March and 2003 May, under observation code BP089.  
Epochs were spaced roughly every
two months, although there were also some significantly
shorter (as short as six days) and longer (as long as six months) spacings between epochs.
Frequencies were chosen to match the observing frequency at which the fastest reported component had been seen
most clearly by J01: 0235+164 was observed at 43 GHz, 0827+243 and 1406$-$076 were observed at 22 GHz.
The total observation time at each epoch was 18 hours
(for the three target sources plus calibrators). This yielded a average rms noise in the uniformly-weighted images
of about 1 mJy beam$^{-1}$, or roughly a factor of four lower than the average rms noise of
4 mJy beam$^{-1}$ for these same three sources in J01.
All observations
recorded dual-circular
polarization.
Only the proper motion results from the total intensity data are reported here; 
flux and polarization variability will be presented
in a subsequent paper.

Calibration and fringe-fitting were done with
the AIPS software package.  Images were produced using standard CLEAN and
self-calibration procedures from the DIFMAP software package.
Naturally weighted images (beam sizes given in Table~\ref{imtab}) of 0235+164, 0827+243, and 1406$-$076
are shown in Figures~1 to 3; the individual source morphologies are discussed in $\S~\ref{results}$.
Natural weighting has been used to better show some of the faint jet components, such as C1 in 0235+164. 
The parameters of these images are given in Table~\ref{imtab}.
Figure~4 shows a set of tapered images of 1406$-$076 (beam sizes given in Figure~4); these tapered images
show the farther out, more extended components in this source. 

\begin{figure}[!ht]
\begin{center}
\includegraphics[scale=0.55]{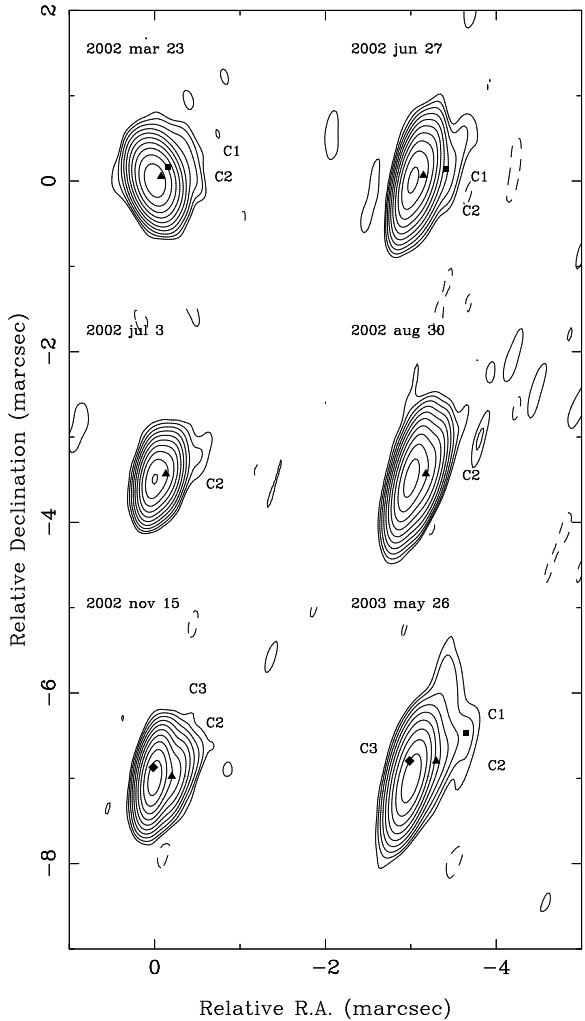}
\end{center}
\caption{
VLBA images of 0235+164.
Image parameters are given in Table~\ref{imtab}.
Model component center positions are shown using the following symbols:
squares for C1, triangles for C2, diamonds for C3.}
\end{figure}

\begin{figure}[!ht]
\begin{center}
\includegraphics[scale=0.55]{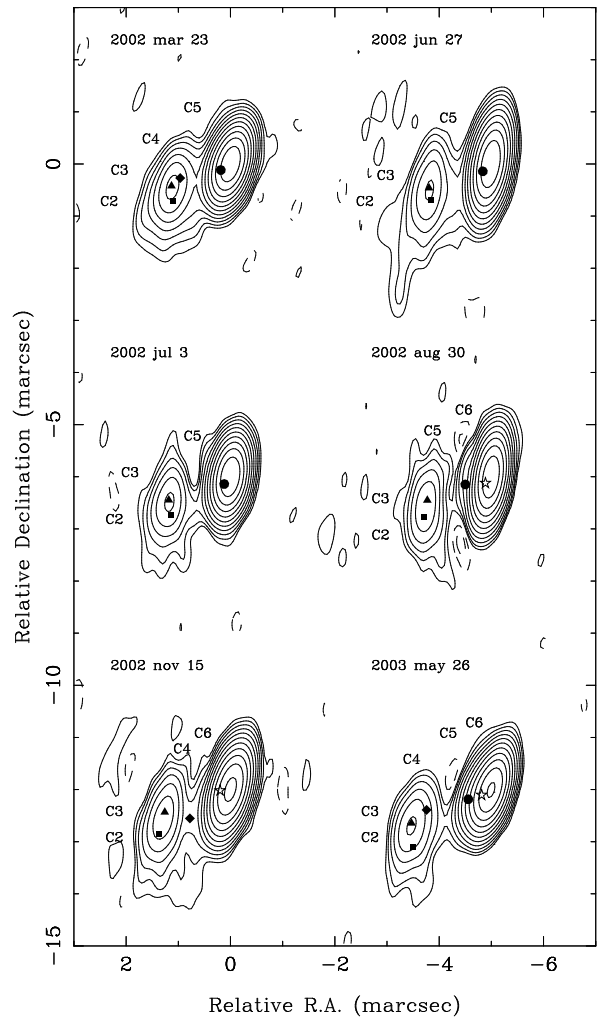}
\end{center}
\caption{
VLBA images of 0827+243.
Image parameters are given in Table~\ref{imtab}.
Model component center positions are shown using the following symbols:
squares for C2, triangles for C3, diamonds for C4, circles for C5, and stars for C6.}
\end{figure}

\begin{figure}[!ht]
\begin{center}
\includegraphics[scale=0.55]{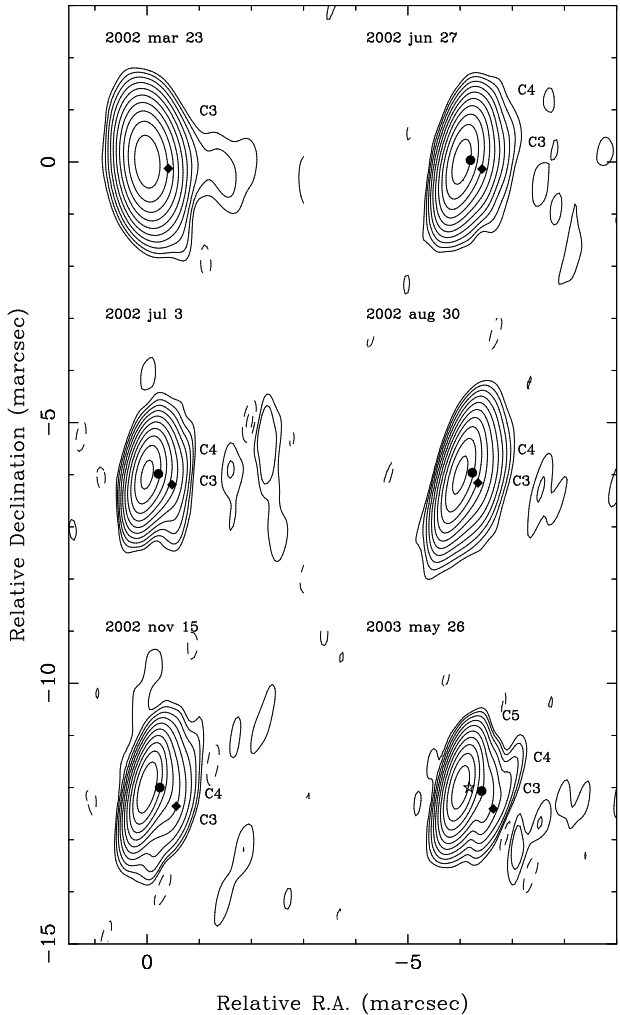}
\end{center}
\caption{
VLBA images of 1406-076.
Image parameters are given in Table~\ref{imtab}.
Model component center positions are shown using the following symbols:
diamonds for C3, circles for C4, and stars for C5.}
\end{figure}

\begin{figure*}[!ht]
\begin{center}
\includegraphics[scale=0.55]{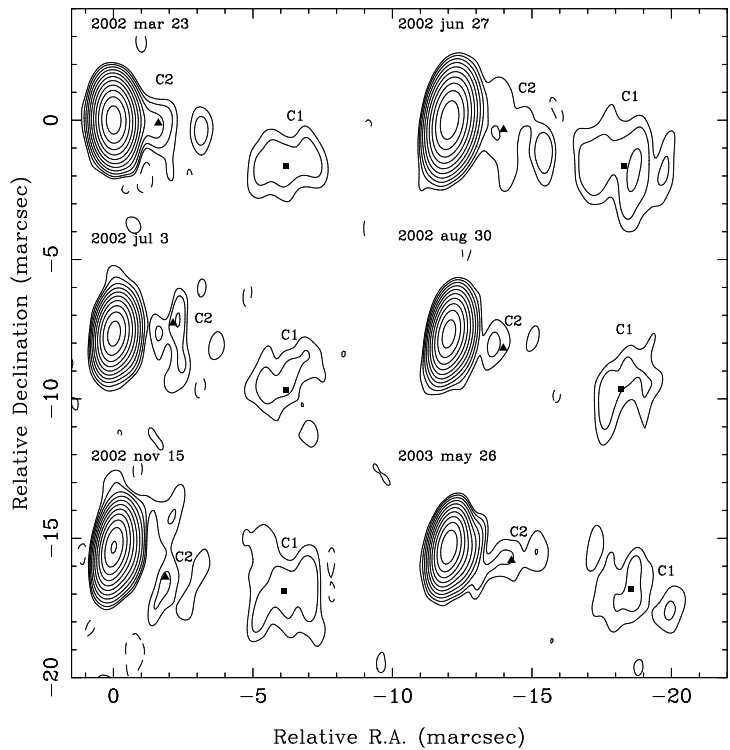}
\end{center}
\caption{
Tapered VLBA images of 1406-076 showing the outer components.
Model component center positions are shown using the following symbols:
squares for C1, and triangles for C2.
Contours are set to the same levels as those in Figure~3.
Beam sizes are 1.32 by 0.63 mas at 1$\arcdeg$, 1.61 by 0.68 mas at $-9\arcdeg$,
1.26 by 0.57 mas at $-7\arcdeg$, 1.37 by 0.54 mas at $-11\arcdeg$,
1.35 by 0.54 mas at $-10\arcdeg$, and 1.3 by 0.53 mas at $-11\arcdeg$.
Peak flux densities are 1.16, 1.02, 0.98, 1.10, 1.13, and 1.05 Jy beam$^{-1}$, respectively.}
\end{figure*}

\begin{table}[!ht]
\caption{Parameters of the Images}
\label{imtab}
\begin{tabular}{l l r c c} \colrule \colrule    
& & & Peak Flux & Lowest \\ 
& & & Density & Contour\tablenotemark{b} \\
Source & Epoch & \multicolumn{1}{c}{Beam\tablenotemark{a}} &
(Jy beam$^{-1}$) & (mJy beam$^{-1}$) \\ \colrule  
0235+164   & 2002 Mar 23 & 0.45,0.23,16.0    & 1.11 & 1.28 \\
           & 2002 Jun 27 & 0.61,0.16,$-$12.2 & 0.83 & 1.36 \\
           & 2002 Jul 3  & 0.43,0.17,$-$10.1 & 0.94 & 1.73 \\
           & 2002 Aug 30 & 0.65,0.17,$-$15.3 & 0.92 & 1.29 \\
           & 2002 Nov 15 & 0.49,0.17,$-$11.2 & 0.85 & 1.07 \\
           & 2003 May 26 & 0.68,0.20,$-$16.5 & 0.79 & 1.90 \\
0827+243   & 2002 Mar 23 & 0.75,0.34,$-$14.1 & 1.16 & 0.63 \\
           & 2002 Jun 27 & 0.93,0.32,$-$8.8  & 1.65 & 1.69 \\ 
           & 2002 Jul 3  & 0.77,0.33,$-$8.1  & 1.63 & 1.64 \\
           & 2002 Aug 30 & 0.88,0.29,$-$7.7  & 1.65 & 1.83 \\
           & 2002 Nov 15 & 0.86,0.32,$-$12.6 & 1.20 & 0.99 \\
           & 2003 May 26 & 0.80,0.32,$-$14.9 & 1.11 & 0.98 \\
1406$-$076 & 2002 Mar 23 & 1.17,0.48,5.8     & 1.11 & 0.76 \\
           & 2002 Jun 27 & 1.15,0.37,$-$12.1 & 0.92 & 1.19 \\
           & 2002 Jul 3  & 0.98,0.32,$-$10.6 & 0.87 & 1.36 \\
           & 2002 Aug 30 & 1.21,0.31,$-$15.1 & 0.98 & 1.47 \\ 
           & 2002 Nov 15 & 1.00,0.30,$-$12.5 & 1.05 & 1.03 \\
           & 2003 May 26 & 1.01,0.29,$-$12.5 & 0.92 & 1.11 \\ \colrule  
\end{tabular}
\tablenotetext{a}{Numbers given for the beam are the FWHMs of the major
and minor axes in mas, and the position angle of the major axis in degrees.
Position angle is measured from north through east.}
\tablenotetext{b}{The lowest contour is set to be three times the rms noise
in the image. Successive contours are each a factor of 2 higher.}
\end{table}

After imaging, Gaussian model components
were fit to the visibility data using the {\em modelfit} task in DIFMAP.
The cores were modeled by elliptical Gaussians, the jet components by circular
Gaussians. (The cores were modeled as circular Gaussians during the early stages of the model-fitting
procedure in order to ensure the correct number of components near the core, at the end of the model-fitting procedure the cores
were allowed to become elliptical Gaussians in order to improve the chi-squared of the fit.)
Model fitting of the visibility data allows sub-beam resolution to be obtained when the signal-to-noise ratio is
high, and the locations of components near the core can be identified accurately in the model fitting even when
these components appear merged with the core on the images.
Identification of separate components near the core is confirmed by preliminary polarization images, which
show changes in polarization properties between the core and the closest component for all three sources.
Parameters of the model fits are given in Table~\ref{mfittab}, and the locations of
model component centers are indicated in Figures~1 to 4.
Components are identified as C1 to C6, from the outermost component inward (too much time
has elapsed between J01's observations and ours to identify any common components, so we do not attempt
to continue J01's naming conventions).
We note the distinction between model `components' (Gaussians fit to the visibilities)
and jet `features' (visually distinct regions on the images).
In particular, asymmetric features may be represented by more than one component;
for example, the bright eastern jet feature in 0827+243 is modeled by Gaussian components C2, C3, and C4,
representing its southern edge, compact center, and trailing edge, respectively.

\begin{table*}[!ht]
\begin{center}
\caption{Gaussian Models}
\label{mfittab}
\begin{tabular}{l l c r r r r r r c} \colrule
& &
& \multicolumn{1}{c}{$S$\tablenotemark{a}} & \multicolumn{1}{c}{$r$\tablenotemark{b}} &
\multicolumn{1}{c}{PA\tablenotemark{b}} &
\multicolumn{1}{c}{$a$\tablenotemark{c}} &
& \multicolumn{1}{c}{$\phi$\tablenotemark{c}} \\
\multicolumn{1}{c}{Source} & \multicolumn{1}{c}{Epoch} & Comp.
& \multicolumn{1}{c}{(mJy)} & \multicolumn{1}{c}{(mas)} &
\multicolumn{1}{c}{(deg)}
& \multicolumn{1}{c}{(mas)} & \multicolumn{1}{c}{$(b/a)$\tablenotemark{c}} & \multicolumn{1}{c}{(deg)}
& $\chi_{R}^{2}$\tablenotemark{d} \\ \colrule
0235+164 & 2002 Mar 22 & Core & 867  & ...   & ...      & 0.07 & 0.62 & 28.7     & 0.49 \\
         &             & C2   & 445  & 0.12  & $-$56.0  & 0.09 & 1.00 & ...      &      \\
         &             & C1   & 35   & 0.25  & $-$46.5  & 0.56 & 1.00 & ...      &      \\
0235+164 & 2002 Jun 27 & Core & 747  & ...   & ...      & 0.09 & 0.28 & 17.7     & 0.55 \\
         &             & C2   & 424  & 0.16  & $-$64.1  & 0.10 & 1.00 & ...      &      \\
         &             & C1   & 16   & 0.43  & $-$71.2  & 0.14 & 1.00 & ...      &      \\
0235+164 & 2002 Jul 3  & Core & 887  & ...   & ...      & 0.10 & 0.42 & 13.3     & 0.56 \\ 
         &             & C2   & 445  & 0.16  & $-$65.1  & 0.10 & 1.00 & ...      &      \\
0235+164 & 2002 Aug 30 & Core & 963  & ...   & ...      & 0.15 & 0.32 & 8.2      & 0.56 \\
         &             & C2   & 333  & 0.19  & $-$73.0  & 0.12 & 1.00 & ...      &      \\
0235+164 & 2002 Nov 15 & Core & 751  & ...   & ...      & 0.11 & 0.37 & 11.2     & 0.66 \\
         &             & C3   & 206  & 0.16  & 6.4      & 0.06 & 1.00 & ...      &      \\
         &             & C2   & 155  & 0.21  & $-$76.3  & 0.15 & 1.00 & ...      &      \\ 
0235+164 & 2003 May 26 & Core & 713  & ...   & ...      & 0.04 & 0.0  & $-$85.5  & 0.50 \\
         &             & C3   & 195  & 0.24  & 5.1      & 0.09 & 1.00 & ...      &      \\
         &             & C2   & 30   & 0.37  & $-$51.1  & 0.10 & 1.00 & ...      &      \\
         &             & C1   & 10   & 0.85  & $-$48.6  & 0.19 & 1.00 & ...      &      \\ 
0827+243 & 2002 Mar 23 & Core & 1139 & ...   & ...      & 0.11 & 0.39 & $-$63.3  & 0.66 \\
         &             & C5   & 123  & 0.23  & 121.6    & 0.13 & 1.00 & ...      &      \\
         &             & C4   & 17   & 1.01  & 105.7    & 0.36 & 1.00 & ...      &      \\
         &             & C3   & 33   & 1.22  & 110.4    & 0.10 & 1.00 & ...      &      \\
         &             & C2   & 29   & 1.33  & 122.7    & 0.52 & 1.00 & ...      &      \\
         &             & C1   & 11   & 5.67  & 135.1    & 1.91 & 1.00 & ...      &      \\
0827+243 & 2002 Jun 27 & Core & 1753 & ...   & ...      & 0.16 & 0.51 & $-$56.0  & 0.65 \\
         &             & C5   & 69   & 0.21  & 133.3    & 0.27 & 1.00 & ...      &      \\
         &             & C3   & 49   & 1.27  & 111.3    & 0.14 & 1.00 & ...      &      \\
         &             & C2   & 52   & 1.34  & 121.0    & 0.73 & 1.00 & ...      &      \\
0827+243 & 2002 Jul 3  & Core & 1669 & ...   & ...      & 0.14 & 0.26 & $-$69.7  & 0.69 \\
         &             & C5   & 126  & 0.19  & 140.3    & 0.22 & 1.00 & ...      &      \\
         &             & C3   & 58   & 1.27  & 110.9    & 0.17 & 1.00 & ...      &      \\
         &             & C2   & 38   & 1.36  & 122.8    & 0.74 & 1.00 & ...      &      \\
0827+243 & 2002 Aug 30 & Core & 1700 & ...   & ...      & 0.15 & 0.06 & $-$89.5  & 0.74 \\
         &             & C6   & 235  & 0.17  & 138.4    & 0.15 & 1.00 & ...      &      \\
         &             & C5   & 12   & 0.53  & 107.0    & 0.0\tablenotemark{e} & ...  & ...      &      \\
         &             & C3   & 52   & 1.31  & 110.4    & 0.18 & 1.00 & ...      &      \\
         &             & C2   & 38   & 1.51  & 121.5    & 0.51 & 1.00 & ...      &      \\
0827+243 & 2002 Nov 15 & Core & 1388 & ...   & ...      & 0.23 & 0.42 & $-$62.7  & 0.87 \\ 
         &             & C6   & 42   & 0.21  & 96.0     & 0.0  & ...  & ...      &      \\
         &             & C4   & 6    & 0.96  & 125.3    & 0.34 & 1.00 & ...      &      \\
         &             & C3   & 53   & 1.34  & 109.0    & 0.25 & 1.00 & ...      &      \\
         &             & C2   & 25   & 1.61  & 122.2    & 0.31 & 1.00 & ...      &      \\ 
0827+243 & 2003 May 26 & Core & 1034 & ...   & ...      & 0.14 & 0.33 & $-$45.7  & 0.61 \\
         &             & C6   & 243  & 0.23  & 119.4    & 0.19 & 1.00 & ...      &      \\
         &             & C5   & 12   & 0.50  & 112.8    & 0.0  & ...  & ...      &      \\
         &             & C4   & 5    & 1.32  & 107.5    & 0.34 & 1.00 & ...      &      \\
         &             & C3   & 40   & 1.68  & 112.8    & 0.24 & 1.00 & ...      &      \\
         &             & C2   & 15   & 1.88  & 126.5    & 0.49 & 1.00 & ...      &      \\
         &             & C1   & 8    & 5.17  & 143.3    & 1.12 & 1.00 & ...      &      \\
1406-076 & 2002 Mar 23 & Core & 1204 & ...   & ...      & 0.22 & 0.0  & $-$86.3  & 0.57 \\
         &             & C3   & 78   & 0.43  & $-$107.7 & 0.0  & ...  & ...      &      \\
         &             & C2   & 7    & 1.62  & $-$93.8  & 0.94 & 1.00 & ...      &      \\
         &             & C1   & 21   & 6.39  & $-$104.8 & 2.01 & 1.00 & ...      &      \\
1406-076 & 2002 Jun 28 & Core & 793  & ...   & ...      & 0.10 & 0.63 & $-$25.3  & 0.54 \\ 
         &             & C4   & 253  & 0.21  & $-$79.8  & 0.16 & 1.00 & ...      &      \\
         &             & C3   & 67   & 0.45  & $-$107.3 & 0.28 & 1.00 & ...      &      \\
         &             & C2   & 12   & 2.02  & $-$99.6  & 2.18 & 1.00 & ...      &      \\
         &             & C1   & 28   & 6.51  & $-$104.7 & 2.38 & 1.00 & ...      &      \\
1406-076 & 2002 Jul 3  & Core & 862  & ...   & ...      & 0.18 & 0.56 & $-$23.8  & 0.65 \\
         &             & C4   & 210  & 0.24  & $-$85.6  & 0.16 & 1.00 & ...      &      \\
         &             & C3   & 48   & 0.53  & $-$110.7 & 0.28 & 1.00 & ...      &      \\
         &             & C2   & 10   & 2.19  & $-$80.0  & 1.34 & 1.00 & ...      &      \\
         &             & C1   & 20   & 6.52  & $-$108.2 & 1.94 & 1.00 & ...      &      \\
1406-076 & 2002 Aug 30 & Core & 986  & ...   & ...      & 0.17 & 0.63 & $-$27.8  & 0.58 \\
         &             & C4   & 157  & 0.25  & $-$78.7  & 0.14 & 1.00 & ...      &      \\
         &             & C3   & 121  & 0.38  & $-$112.4 & 0.38 & 1.00 & ...      &      \\
\end{tabular}
\end{center}
\end{table*}

\begin{table*}[!ht]
\begin{center}
TABLE 2---{\em Continued} \\
\begin{tabular}{l l c r r r r r r c} \colrule
& &
& \multicolumn{1}{c}{$S$\tablenotemark{a}} & \multicolumn{1}{c}{$r$\tablenotemark{b}} &
\multicolumn{1}{c}{PA\tablenotemark{b}} &
\multicolumn{1}{c}{$a$\tablenotemark{c}} &
& \multicolumn{1}{c}{$\phi$\tablenotemark{c}} \\
\multicolumn{1}{c}{Source} & \multicolumn{1}{c}{Epoch} & Comp.
& \multicolumn{1}{c}{(mJy)} & \multicolumn{1}{c}{(mas)} &
\multicolumn{1}{c}{(deg)}
& \multicolumn{1}{c}{(mas)} & \multicolumn{1}{c}{$(b/a)$\tablenotemark{c}} & \multicolumn{1}{c}{(deg)}
& $\chi_{R}^{2}$\tablenotemark{d} \\ \colrule
         &             & C2   & 9    & 2.05  & $-$104.4 & 1.35 & 1.00 & ...      &      \\
         &             & C1   & 23   & 6.50  & $-$107.5 & 2.05 & 1.00 & ...      &      \\
1406-076 & 2002 Nov 15 & Core & 1035 & ...   & ...      & 0.13 & 0.24 & $-$46.4  & 0.63 \\
         &             & C4   & 186  & 0.25  & $-$89.3  & 0.31 & 1.00 & ...      &      \\
         &             & C3   & 29   & 0.67  & $-$122.1 & 0.17 & 1.00 & ...      &      \\
         &             & C2   & 4    & 2.15  & $-$119.9 & 0.0  & ...  & ...      &      \\
         &             & C1   & 21   & 6.33  & $-$104.3 & 2.03 & 1.00 & ...      &      \\
1406-076 & 2003 May 27 & Core & 966  & ...   & ...      & 0.19 & 0.0  & $-$59.2  & 0.63 \\
	 &             & C5   & 131  & 0.18  & $-$91.9  & 0.0  & ...  & ...      &      \\
         &             & C4   & 85   & 0.43  & $-$99.3  & 0.31 & 1.00 & ...      &      \\
         &             & C3   & 24   & 0.76  & $-$122.6 & 0.21 & 1.00 & ...      &      \\
         &             & C2   & 12   & 2.32  & $-$101.9 & 1.74 & 1.00 & ...      &      \\
         &             & C1   & 20   & 6.72  & $-$102.9 & 1.98 & 1.00 & ...      &      \\ \colrule  
\end{tabular}
\end{center}
$^{\rm{a}}$ Flux density of the component.\\
$^{\rm{b}}$ $r$ and PA are the polar coordinates of the
Gaussian center relative to the core.
PA is measured from north through east.\\
$^{\rm{c}}$ $a$ and $b$ are the FWHM of the major and minor axes of the Gaussian,
and
$\phi$ is the position angle of the major axis.\\
$^{\rm{d}}$ Reduced Chi-squared of the model.\\
$^{\rm{e}}$ The size of a component occasionally goes to zero in DIFMAP model fitting,
if the size is not well constrained by the data.\\
\end{table*}

\section {Results}
\label{results}
Apparent speeds for the jet components in Table~\ref{mfittab} 
were derived 
from linear least-squares fits to the separation of the components from the VLBI core
versus time at two or more epochs
(with the exception of the diffuse component C1 in 0827+243).
These fits are shown in Figure~5.
Component C1 in 1406$-$076 is considerably farther from the core (separation about 6.5 mas) than the other components,
so is not shown in Figure~5.
The disappearance of a component on Figure~5 followed by its re-appearance at later epochs
(e.g., component C1 in 0235+164) does not imply the literal disappearance and re-appearance of the 
component on the images.  Rather, the component is usually present at a marginally significant level
in the intervening images, but is not significant enough to have its properties well constrained by the 
model fitting.  Multiple components close to the core are occasionally blended into a single component by model fitting,
e.g., the first three detections of `C5' in 0827+243 are most likely a blend of components C5 and C6 that
cannot be separated by the resolution of these observations.

\begin{figure}[!ht]
\begin{center}
\includegraphics[scale=0.55]{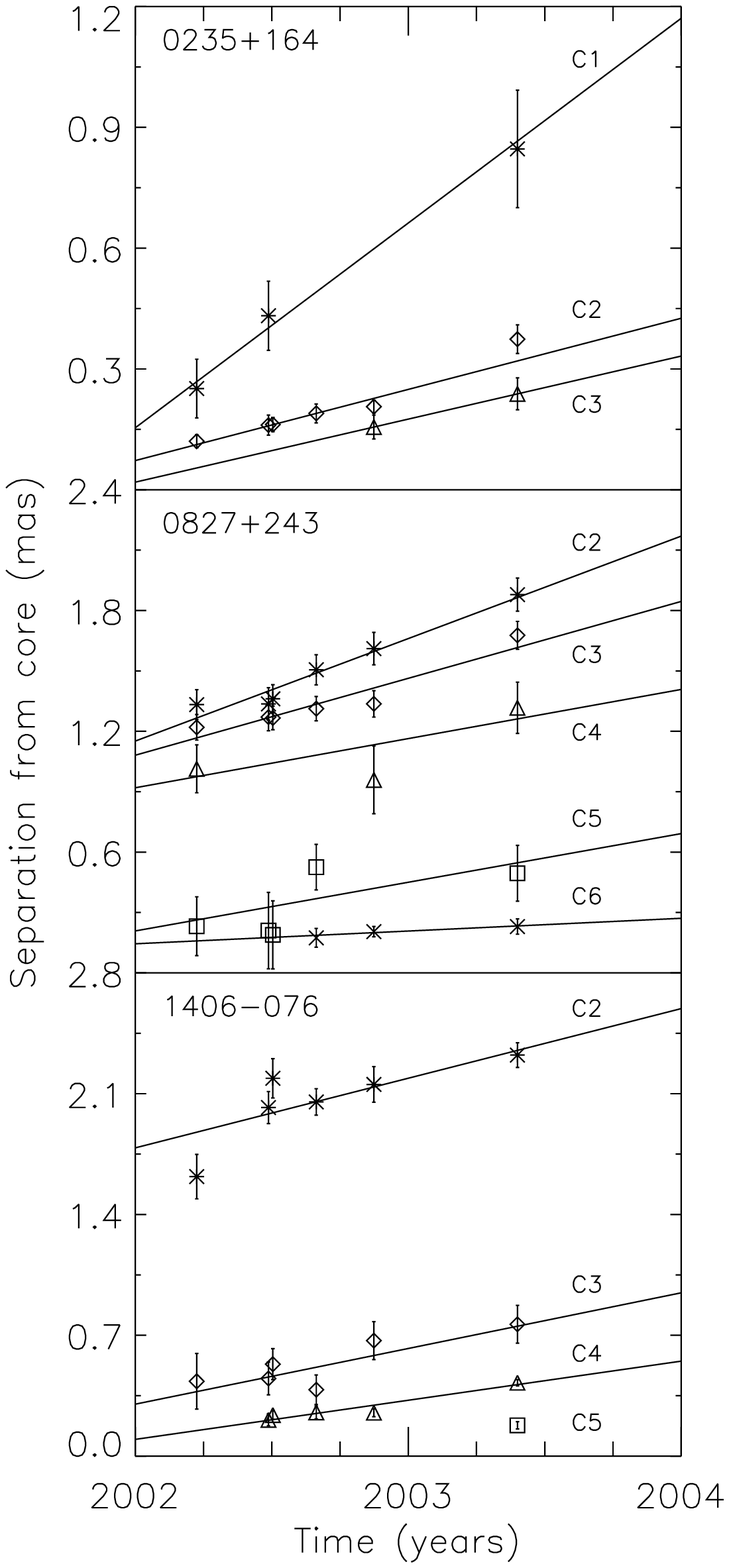}
\end{center}
\vspace{-0.20in}
\caption{Distances from the core of Gaussian component centers as
a function of time.
The lines are the least-squares fits to outward motion with constant speed.}
\end{figure}

Error bars on component separations were defined to be a fraction $1/2^{n}$ of the beam size
\footnote{Maximum extent of the projection of the
two-dimensional beam onto a line joining the center of the core to the center of the Gaussian component.},
where $n$ was an integer from 1 to 4 that was set according to the average dynamic range of the component detection
(see Table~\ref{speedtab}).
The relation between $n$ and the dynamic range was calibrated  using the scatter in the model-fit
positions between components in the second and third epochs, which were only six days apart
(hence differences in model-fit positions between these two epochs are due to errors rather
than component motions). 
We note that the fitted apparent speeds and associated errors are not significantly different
if no error bars are specified for the component positions, i.e., if the error in the apparent speed is determined
by the scatter of the positions about the fit, as in Kellermann et al. (2004) (hereafter K04).

The measured apparent speeds are tabulated in Table~\ref{speedtab}.
A ``quality code'' (Excellent, Good, Fair, or Poor) is assigned to each 
apparent speed measurement in Table~\ref{speedtab},
using the same criteria used for the 2~cm survey
(see K04 for definitions
of the quality codes).
The peak measured apparent speeds were $25.6\pm7.0~c$, $25.6\pm4.4~c$, and $28.2\pm6.6~c$
for 0235+164, 0827+243, and 1406$-$076, respectively.
The minimum Lorentz factor that can produce an apparent speed, $\beta_{app}$,
is determined as $\Gamma_{min}=\sqrt{1+\beta_{app}^2}$, where $\beta_{app}$
is in units of the speed of light. For the highest apparent speed in each
source, $\Gamma_{min}$ is approximately equal to the apparent speed
(assuming these fastest measured apparent pattern speeds are equal to or less than the apparent bulk speed).
Corresponding maximum angles to the line-of-sight ($\theta_{max}=2\arctan(1/\beta_{app}$))
are $\theta_{max}=$ 4.5, 4.5, and 4.1 degrees for 0235+164, 0827+243, and 1406$-$076, respectively.
We measure significantly different apparent speeds for the different components in the jets of 0235+164 and 0827+243;
this detection of different apparent speeds for different components has a significance exceeding 3$\sigma$
for 0827+243.
Implications of these different apparent speeds are discussed in $\S~\ref{discussion}$.

\begin{table}[!ht]
\caption{Apparent Speeds}
\label{speedtab}
\begin{tabular}{l c c c c c} \colrule
& & $\mu$ \\
\multicolumn{1}{c}{Source} & Comp. & (mas yr$^{-1}$) & $\beta_{app}$
& $n$\tablenotemark{a} & Quality\tablenotemark{b} \\ \colrule
0235+164 & C1 & $0.51\pm0.14$ & $25.6\pm7.0$  & 2 & P \\
         & C2 & $0.18\pm0.03$ & $8.9\pm1.3$   & 4 & G \\
         & C3 & $0.16\pm0.09$ & $7.9\pm4.7$   & 4 & P \\
0827+243 & C2 & $0.51\pm0.09$ & $25.6\pm4.4$  & 3 & E \\
         & C3 & $0.38\pm0.07$ & $19.2\pm3.7$  & 3 & E \\
         & C4 & $0.24\pm0.15$ & $12.3\pm7.4$  & 2 & F \\
         & C5 & $0.24\pm0.16$ & $12.1\pm8.1$  & 2 & P \\
         & C6 & $0.06\pm0.07$ & $3.2\pm3.7$   & 4 & F \\
1406-076 & C1 & $0.22\pm0.19$ & $15.6\pm13.2$ & 1 & P \\
         & C2 & $0.40\pm0.09$ & $28.2\pm6.6$  & 2 & F \\
         & C3 & $0.32\pm0.13$ & $22.5\pm8.9$  & 2 & F \\
         & C4 & $0.23\pm0.03$ & $15.8\pm2.0$  & 4 & G \\ \colrule  
\end{tabular}
\tablenotetext{a}{Radial error bars are expressed as a fraction $1/2^{n}$ of the beam.}
\tablenotetext{b}{Quality code: Excellent (E), Good (G), Fair (F), or Poor (P).}
\end{table}

If non-radial motion is present, then the apparent speeds in Table~\ref{speedtab} will be lower limits to the
actual apparent speed (i.e., they will represent only the radial component).
We investigated possible non-radial motions using the method described by K04. 
Separate linear fits were made to $x(t)$ and $y(t)$; this provides a fit to the direction of motion.
A direction of motion significantly different from the average position angle indicates non-radial motion.
We found 
significant non-radial motion ($>3\sigma$, toward the south) only for C2 in 1406$-$076.
The possibility of accelerated radial motion was also investigated, by performing second-order fits to $r(t)$,
and checking the significance of the second-order term.
Radial acceleration at borderline significance (2.2$\sigma$) was found only for C4 in 1406$-$076.
While these deviations from ballistic motion in 1406$-$076 are formally significant,
C2 is a faint (9 mJy) component that received only a `Fair' quality code, and the direction
of its non-radial motion is in the direction of poor angular resolution for this
low-declination source.
Thus, while it seems qualitatively clear that C2 exhibits non-radial motion toward the south, 
we have not replaced the linear fits in Table~\ref{speedtab} with the nonlinear values.

Below we discuss some of the specific results for individual sources:

0235+164.---($z=0.94$) This source has been noted as having 
varying component position angles since its early VLBI observations
(Jones et al. 1984).  We confirm the morphology of this source seen by J01 ---
the source has components at varying position angles:
to the northwest at about $-55\arcdeg$ (our C1 and C2, and J01's B1),
and to the north at about $5\arcdeg$ (our C3 and J01's B2).
As noted by J01, this indicates a broad projected opening angle ($\sim 60\arcdeg$)
within about 0.5 mas of the core.  The peak apparent speed
measured in this paper ($25.6\pm7.0~c$ for C2)
is different at the 2$\sigma$ level from the peak apparent speed measured by J01 ($46.5\pm8.0~c$ for their B1).
A Lorentz factor equal to
our $\Gamma_{min}$ of 26 for this source would allow Doppler
factors up to 52.
Very high Doppler factors of $\delta\sim100$ have been suggested for this source, based on its
radio variability (Kraus et al. 1999) and high 
brightness temperature (Frey et al. 2000).
K04 classify 0235+164 as a naked core at 15 GHz, suggesting that the components in this
source are not long lived and are detectable only close to the core.

0827+243.---($z=0.939$) Of the fastest components in our three sources, only C2 in 0827+243
has an `Excellent' quality code.  This is because the bright eastern jet feature 
(made up of Gaussian components C2, C3, and C4) is a
distinct feature that is not merged with the core or any other features, so that there is no
ambiguity in tracking the feature from epoch to epoch.  
Figure~6 shows a mosaic of four of our
six images of 0827+243, where the motion of this feature at $\sim25~c$
is clearly visible.  
The proper motion that we measure for component C2 agrees well with the proper
motion of J01's component D2 (but note that these are not the same component), 
the difference in the apparent speeds for these components between this paper
and J01 is due solely to the different redshifts used in the two papers
(see footnote (1) of Jorstad \& Marscher, 2004).  
Analysis of the large-scale jet (Jorstad \& Marscher 2004) 
has shown it to be highly relativistic out to kiloparsec scales.

\begin{figure}[!ht]
\begin{center}
\includegraphics[scale=0.45]{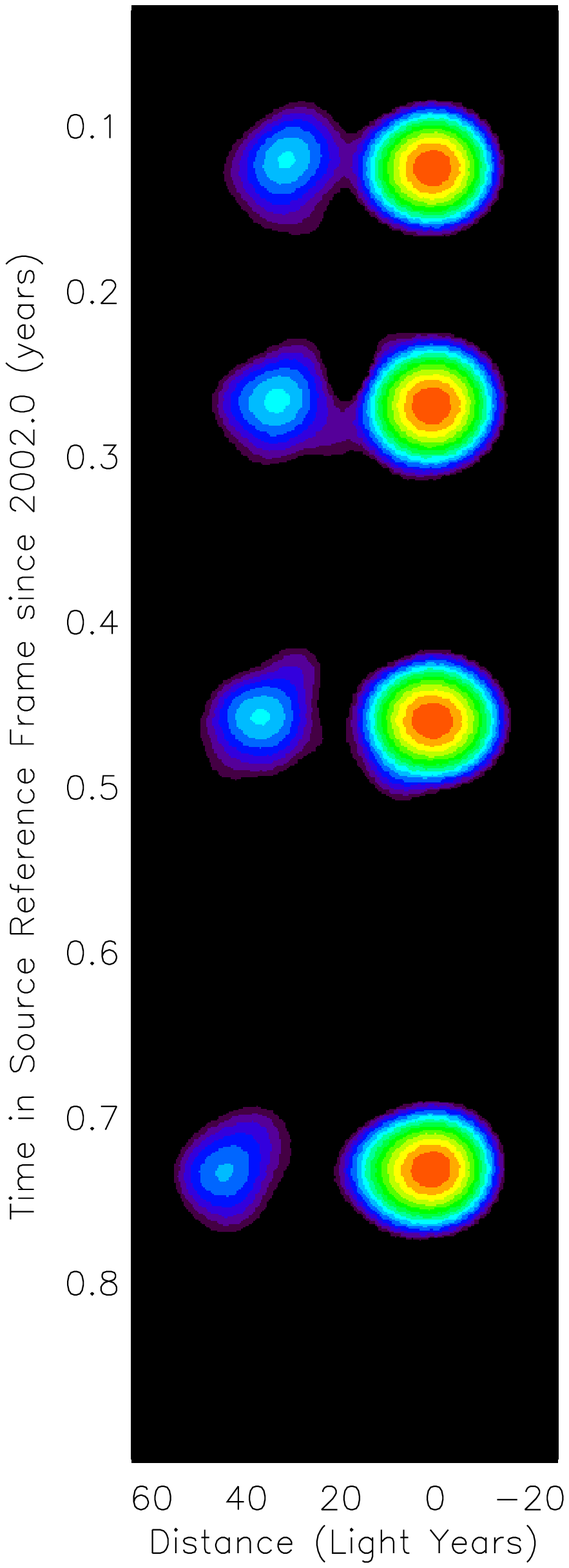}
\end{center}
\vspace{0.15in}
\caption{Mosaic of images of 0827+243 at 22 GHz.
The bright feature moves approximately 15 light-years in 0.6 years (source-frame),
for an apparent speed of about $25~c$.  Only four of the six epochs are shown to prevent
overlapping of images.
The peak flux densities at the four epochs are 1.2, 1.6, 1.2, and 1.1 Jy beam$^{-1}$, respectively.
Images have been rotated 25$^{\circ}$ clockwise, and restored with a circular 0.5 mas beam.
Model component C3 is at the center of the bright jet feature.}
\end{figure}

1406$-$076.---($z=1.49$)
We find a similar morphology to J01.
The jet dims rapidly with distance from the core, and the components become faint and
difficult to follow at 22 GHz.
Non-radial motion of a component toward the south was seen by J01 for their component B3,
similar to the path followed by our C2.  However, the fitted speeds of those components are
different at the 2$\sigma$ level: J01 measure $43.8\pm3.5~c$ for their B3, we measure 
$28.2\pm6.6~c$ for our C2.  Note also that the very high apparent speed we measure for C2
is in part due to the low value of the radial position at the first epoch (see Figure~5,
without this first point the fitted apparent speed is a somewhat lower $20.4\pm7.4~c$),
although this position seems well-determined (see Figure~4).

We conclude this section by noting that our measurements of 
apparent speeds exceeding 25~$c$ in 0235+164 and 1406$-$076 are, like the
measurements of the apparent speeds of those two sources in J01, 
only of marginal confidence.  This is not through any fault of the data, since
these are high-sensitivity full-track VLBA observations, but is due to the fact that the
jets in these two sources fade rapidly with distance from the core, so that components become faint and difficult
to follow. A very fast apparent speed for component C2
in 0827+243, however, has been established with a high degree of confidence.

\section{Discussion}
\label{discussion}
Inspection of Table~\ref{speedtab} shows that different Gaussian components in the
same source have different apparent speeds.  There are two possible origins of
these different apparent speeds.  The first possibility is that the components move with different
pattern speeds that are not necessarily equal to the bulk speed.
Lister (2005) concluded from a correlation of apparent speeds from the 2~cm and MOJAVE surveys
with other source properties
that various pattern speeds are present in the jets, but that the fastest measured
pattern speed is approximately equal to the bulk speed (i.e., patterns are formed
that move with speeds up to and including the bulk speed).
In that case, our fastest measured pattern speeds of $\sim 25-30~c$ would be the speeds
most indicative of the bulk apparent jet speeds for these sources in 2002 and 2003.
The differences between our peak apparent speeds and those measured by J01 for 
0235+164 and 1406$-$076 of $\sim45~c$ could then be due to either an actual
decrease in the bulk Lorentz factor of the flow between 1995 and 2002, or the fact that no component
with a pattern speed quite equal to the higher bulk speed of the jet existed during the
time of our monitoring.
An alternative to different bulk and pattern speeds is that the apparent bulk speed of
the components changes with distance from the core, due to either a change
in bulk Lorentz factor or viewing angle.  This is suggested by the increase of apparent speed
with distance from the core in all of our sources, and is expected in some models
(e.g., Vlahakis \& K\"{o}nigl 2004).  However, such changes would also yield predictable changes in the
brightness of components due to changing Doppler boosting, and these changes are not seen
(however, such changes would be complicated by the intrinsic changes in the brightness of components). 

A consensus appears to be forming from several recent studies
that peak apparent speeds in blazar jets are about $30~c$,
and therefore that bulk Lorentz range up to $\Gamma\sim30$.
Example peak apparent speeds include $32.8~c$ from K04, 
$26.8~c$ from Jorstad et al. (2005) (excluding 1510$-$089, see below),
and $32.5~c$ from Piner et al. (2004). 
While there is general agreement on the statistical distribution of apparent speeds,
there is considerable disagreement for apparent speeds of individual sources 
(e.g., 0528+134 in K04 and Jorstad et al. (2005)).
Whether such different measurements for the same sources
reflect actual differences in the jet with time or observing frequency,
or whether they are related to different component identifications by 
different observers, remains to be seen.
The two current outliers to the apparent speed distribution, with apparent speeds well in excess
of $30~c$, are 1502+106, $49.7\pm12.4~c$ (An et al. 2004),
and 1510$-$089, 
$45.6\pm3.6~c$ (Jorstad et al. 2005).  Both sources merit continued study, to
see if $\Gamma\sim50$ components are repeatedly produced in these jets.

These measurements of peak Lorentz factors attained by parsec-scale jets are
useful in providing a value that must be reached by jet acceleration models.
Current numerical simulations do not produce jets with Lorentz factors
exceeding $\Gamma\sim2$, but this is most likely due to the limited acceleration region
caused by the limits to the numerical grid size (Koide 2004).
Analytical models that are not constrained by these size limits
(but that make simplifying
assumptions to make the analysis tractable)
have been successful in predicting 
acceleration of components to $\Gamma\approx35$ (Vlahakis \& K\"{o}nigl 2004).
The interplay between the apparent speed observations and the theoretical models should continue to be
a valuable tool for constraining acceleration mechanisms of relativistic jets.

\acknowledgments
Part of the work described in this paper has been carried out at the Jet
Propulsion Laboratory, California Institute of Technology, under
contract with the National Aeronautics and Space Administration.
We acknowledge the constructive comments given by the anonymous referee.
This work was supported by the
National Science Foundation under Grant No. 0305475,
and by a Cottrell College Science Award from Research Corporation.

\end{document}